\documentclass[aps,prl,twocolumn,showpacs,floatfix,amsmath,amssymb]{revtex4-2}

\usepackage{placeins}

\usepackage[colorlinks=true,linkcolor=blue,citecolor=blue,urlcolor=blue]{hyperref}
\usepackage{graphicx} 
\usepackage{soul} 

\begin{document}

\title{Light-Induced Transient Polarization Reversal in Rhombohedrally Stacked Bilayer Transition Metal Dichalcogenides via an Electronic Mechanism}

\author{Xiangzhou Zhu} \email{xiangzhou.zhu@unitn.it}

\author{Stefano Mocatti}

\author{Matteo Calandra} \email{m.calandrabuonaura@unitn.it}

\affiliation{{Department of Physics}, {University of Trento}, {Via Sommarive 14}, {38123} {Povo}, {Italy}}

\begin{abstract}
Light-induced sliding ferroelectricity in two-dimensional van der Waals materials enables polarization control via relative layer motion. However, polarization switching occurs on the time scale of shear modes (tens of ps) and requires very large fluences, potentially damaging the samples. Here, using constrained density functional theory and many-body real-time simulations, we demonstrate an ultrafast electronic reversal of the total out-of-plane polarization sign in the photoexcited state, without requiring interlayer sliding, in rhombohedrally stacked transition-metal dichalcogenide bilayers. The polarization changes sign relative to its initial ground-state value at moderate fluences and within $\sim$ 200 fs, about 50 times faster than the typical shear-mode period. The ultrafast switching is driven by a rearrangement of localized dipoles around the tungsten sites and can be probed by phase sensitive transient second harmonic generation measurements. We establish a novel general mechanism for electronic control of low-dimensional ferroelectrics common to all polar multilayers having type II band alignment. Our work has
direct implications for ultrahigh-speed volatile optical memory operating on sub-ps time scales.
\end{abstract}

\maketitle
Ferroelectric materials, whose spontaneous polarization can be reversed by external fields, are promising platforms for next-generation memory and electronic technologies \cite{arimotoCurrentStatusFerroelectric2004,scottApplicationsModernFerroelectrics2007,khanFutureFerroelectricFieldeffect2020}.
While conventional bulk ferroelectrics enable robust polarization control, their integration into nanoscale devices is hindered by strong depolarization fields at reduced thickness \cite{batraThermodynamicStabilityThin1972,mehtaDepolarizationFieldsThin1973}. 
Moreover, polarization switching relies on slow ionic and domain-wall motion, fundamentally limiting the achievable switching speed \cite{merzDomainFormationDomain1954,arimotoCurrentStatusFerroelectric2004,nelsonDomainDynamicsFerroelectric2011}.
As modern electronics increasingly demand ultrathin, fast-switching components, there is a growing need for alternative ferroelectric mechanisms that operate efficiently in the few-layer regime.

\begin{figure}[ht]
    \centering
    \includegraphics[width=\linewidth]{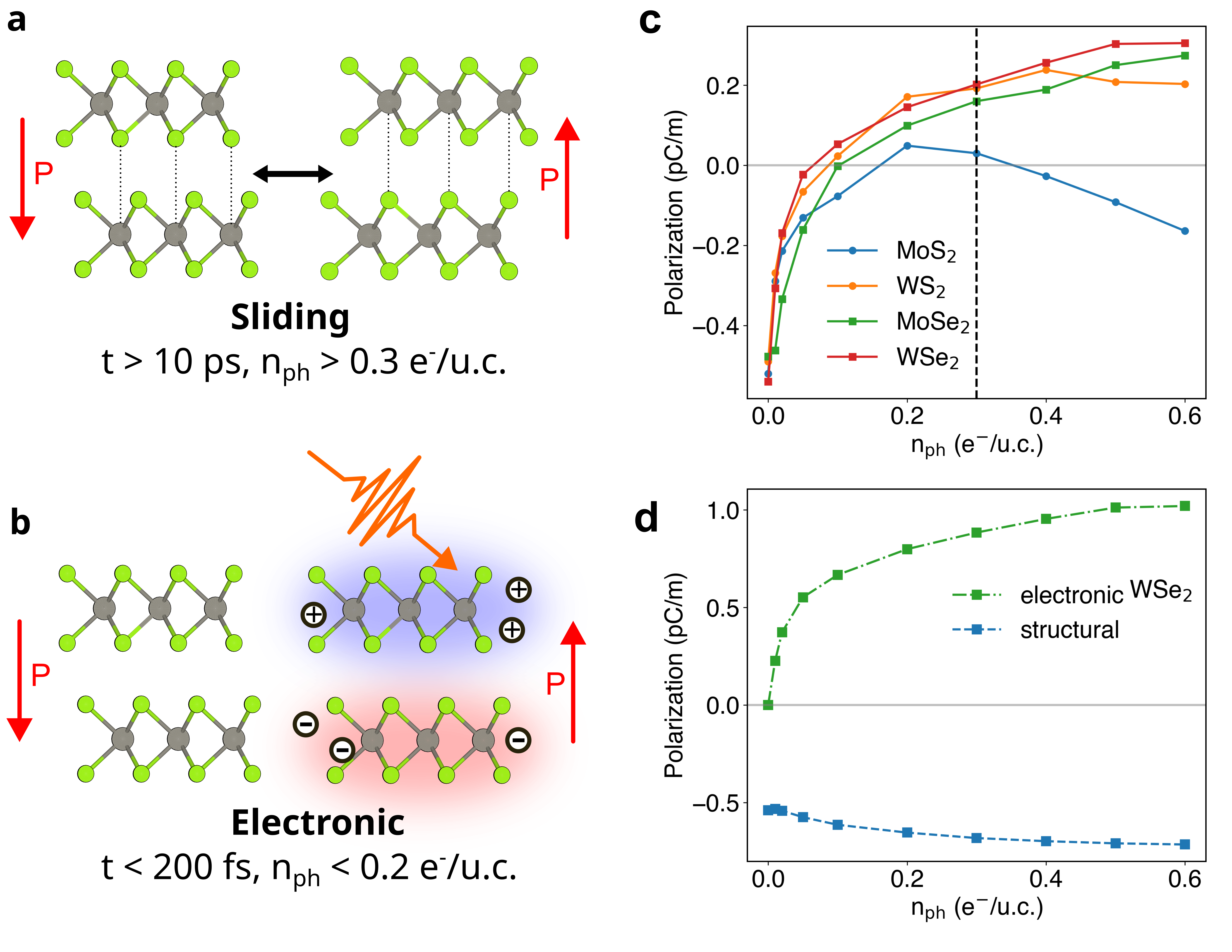}
    \caption{\textbf{a}. Schematic illustration of sliding ferroelectricity in rhombohedrally stacked bilayers.
    Here, M and X denote the transition-metal and chalcogen atomic planes, respectively. The R(MX) stacking (left) and R(XM) stacking (right) correspond to opposite out-of-plane polarization directions. Polarization reversal is achieved by optically induced in-plane sliding of one layer relative to the other, on time scales typical of shear modes. \textbf{b}. Electronically driven polarization switching induced by ultrafast optical charge transfer from the bottom layer to the top one. This mechanism does not require atomic motion and occurs on the fs time scale at much lower fluences than sliding ferroelectric switching. 
    \textbf{c}. Polarization as a function of photoexcited carrier density $n_\mathrm{ph}$ for MoS$_2$ (blue), WS$_2$ (orange), MoSe$_2$ (green), and WSe$_2$ (red). The dashed line indicates the onset carrier density of the phonon instability leading to structural distortion in WS$_2$, MoSe$_2$, and WSe$_2$.
\textbf{d}. Structural (blue) and electronic (green) contributions to the WSe$_2$ polarization as a function of photoexcited carrier density. }
    \label{fig:fig1}
\end{figure}

Two-dimensional van der Waals ferroelectrics have recently emerged as a possible solution, as they can sustain switchable polarization down to the monolayer limit \cite{zhangFerroelectricOrderVan2022,wangTwodimensionalVanWaals2023,liVanWaalsFerroelectrics2024}.
Among these systems, sliding ferroelectrics have attracted significant interest \cite{liBinaryCompoundBilayer2017,wuSlidingFerroelectricity2D2021,wang2DVanWaals2025}.
In these materials, out-of-plane polarization arises from symmetry breaking in specific stacking orders and can be tuned by in-plane sliding, as illustrated in Fig.~\ref{fig:fig1}(a).
This behavior has been experimentally observed in 1T-WTe$_2$ \cite{feiFerroelectricSwitchingTwodimensional2018}, h-BN \cite{viznersternInterfacialFerroelectricityVan2021,yasudaStackingengineeredFerroelectricityBilayer2021}, and 3R-stacked TMD bilayers \cite{wangInterfacialFerroelectricityRhombohedralstacked2022,debCumulativePolarizationConductive2022,westonInterfacialFerroelectricityMarginally2022}, and shows potential for applications in non-volatile memory \cite{yasudaUltrafastHighenduranceMemory2024,yangFerroelectricTransistorsBased2024,liSlidingFerroelectricMemories2024,chenRoomtemperatureMultiferroicitySliding2025}, ferroelectric tunnel junctions (FTJs) \cite{wanRoomTemperatureFerroelectricity12022,gaoTunnelJunctionsBased2024}, and neuromorphic computing \cite{liSlidingFerroelectricMemories2024,zhaoUltrafastLightModulatedSliding2025}.

Optical excitation is widely used to manipulate electronic structure and lattice degrees of freedom in 2D van der Waals materials \cite{baoLightinducedEmergentPhenomena2022}.
Recent studies \cite{yangLightInducedCompleteReversal2024,gaoLargePhotoinducedTuning2024,wangUltrafastSwitchingSliding2024} have shown that ultrafast laser pulses can tune the polarization of sliding ferroelectrics, but only at very large photoexcited carrier densities and through structural distortions, with switching on phonon time scales ($\sim$10 ps) \cite{sieUltrafastSymmetrySwitch2019,fukudaUltrafastDynamicsLow2020,gaoLargePhotoinducedTuning2024}. 
However, achieving such large photocarrier densities requires high fluences, which can generate substantial heat in the systems.
Thus, it would be highly desirable to achieve optical switching at lower fluences and on time scales comparable to photocarrier relaxation, i.e., a few hundred femtoseconds.
Establishing an ultrafast and electronically driven switching pathway would enable volatile optical memory architectures, with light-controlled operation at speeds not achievable with structurally driven ferroelectric mechanisms.

Motivated by these questions, in this Letter we investigate the light-induced polarization dynamics of rhombohedrally stacked bilayer transition-metal dichalcogenides (TMDs). By using constrained density-functional theory (cDFT) \cite{mariniLatticeDynamicsPhotoexcited2021} to assess how the photoexcited carrier density affects the ferroelectric (FE) polarization, we find that WSe$_2$ exhibits the lowest switching threshold: the total polarization changes sign at a moderate carrier density of 0.1 e$^{-}$/u.c., well below that required to activate shear phonon modes ($>$ 0.3 e$^{-}$/u.c.). Using many-body real-time simulations of photocarrier relaxation \cite{Mocatti_2026}, we show that once this threshold is exceeded, the total polarization changes sign relative to its initial ground-state value in about 200 fs, a timescale $\approx 50$ times shorter than the sliding-induced structural response on the ps scale. Combining polarized band analysis with time-dependent charge redistribution, we find that this sign reversal is driven by interlayer charge transfer, as schematically shown in Fig.~\ref{fig:fig1}b. 
Because this mechanism is driven by photoexcited charge redistribution rather than a structural transition, it corresponds to a transient reversal of the total polarization in the photoexcited state, rather than permanent switching between ferroelectric states induced by interlayer sliding (shown schematically in Fig.\ref{fig:fig1}a). 
The reversed polarization is therefore expected to persist only while the nonequilibrium carrier imbalance survives and to recover its initial value after carrier recombination. 
These results identify a microscopic mechanism for ultrafast electronic control of 2D ferroelectrics and point to their potential use in volatile, high-speed optoelectronic memory devices. Further technical details, including explanations on how to detect the polarization inversion in the transient state, are provided in the Supplemental Material (SM) \cite{SupplementalMaterial}.

We begin by quantifying how the out-of-plane polarization of rhombohedrally stacked TMD bilayers responds to ultrafast light excitation. 
Using cDFT (see SM for additional technical details \cite{SupplementalMaterial}), we compute the polarization as a function of the photoexcited carrier density $n_{\text{ph}}$. In this approach, two independent Fermi levels are imposed for electrons and holes to describe the quasi-equilibrium thermalized electron-hole plasma after carrier thermalization and before electron-hole recombination. This justifies the use of cDFT to analyze the carrier-density dependence of the polarization, while the relaxation timescale is established independently by the real-time simulations.

The results of our calculations are displayed in Fig.~\ref{fig:fig1}c. For all four materials, the polarization magnitude decreases sharply and crosses zero within 0.05–0.2 e$^{-}$/u.c., showing that moderate photoexcitation is sufficient to overcome the initial intrinsic polarization.
MoS$_2$ alone displays a nonmonotonic behavior, in which the polarization becomes positive and then returns to negative at higher carrier densities. 
This behavior arises from its distinct band alignment (see SM section II \cite{SupplementalMaterial}) and was reported previously.
Among the four systems, WSe$_2$ exhibits the lowest switching threshold, with polarization reversal already below 0.1 e$^{-}$/u.c.
Importantly, this sign reversal occurs below the onset of the phonon-driven structural instability marked by the dashed line in Fig.~1c (see Fig.~S4 in the SM \cite{SupplementalMaterial}), showing that the switching sets in while the lattice remains dynamically stable. We therefore use WSe$_2$ as the model material for further detailed analysis.

To identify the origin of this reversal, Fig.~\ref{fig:fig1}d decomposes the polarization of WSe$_2$ into structural and electronic contributions.
Let $\mathbf{R}(n_\mathrm{ph})$ denote the lattice geometry relaxed in the presence of a photoexcited carrier density $n_\mathrm{ph}$. The total polarization can then be written as $P_{\mathrm{tot}}(n_\mathrm{ph})=P(n_\mathrm{ph},\mathbf{R}(n_\mathrm{ph}))$. To isolate the effect of lattice relaxation alone, we evaluate the polarization on the same relaxed geometry after removing the photoexcited carriers, $P_{\mathrm{struct}}(n_\mathrm{ph})=P(0,\mathbf{R}(n_\mathrm{ph}))$. This quantity measures the change in polarization arising solely from the photoinduced lattice distortion and varies by less than 0.2 pC/m (blue line). The remaining part, $P_{\mathrm{el}}(n_\mathrm{ph})=P_{\mathrm{tot}}(n_\mathrm{ph})-P_{\mathrm{struct}}(n_\mathrm{ph})$, defines the electronic contribution, which increases by more than 1 pC/m (green line) and dominates the sign reversal. This shows that the observed sign reversal is driven primarily by the redistribution of photoexcited carriers rather than by photoinduced lattice relaxation.

Notably, the structural and electronic contributions have opposite signs, indicating that lattice relaxation induces a dipole that partially counteracts the light-induced electronic dipole.
The dominance of the electronic contribution shows that the polarization reversal originates from photoinduced charge redistribution rather than structural distortions.
We stress that the structural contribution discussed here does not correspond to switching between the two stacking configurations shown in Fig.~1a; rather, it reflects a sub-ps structural photoinduced relaxation without symmetry breaking.

At very high photocarrier densities, a different mechanism emerges, namely, the system becomes structurally unstable,
forming two lattice-distorted phases with opposite polarizations, as reported previously \cite{gaoLargePhotoinducedTuning2024}.
For all considered selenides and for WS$_2$, the structural instability occurs around 0.3 e$^{-}$/u.c., as indicated by the black dashed line in Fig.~\ref{fig:fig1}c, while for MoS$_2$ it occurs at much higher fluences, namely near 0.9 e$^{-}$/u.c (see SM Fig.~S4 \cite{SupplementalMaterial}). 
These structural transitions occur only at much larger excitation densities than those required for the electronically driven polarization inversion discussed here.

\begin{figure}[ht]
    \centering
    \includegraphics[width=0.9\linewidth]{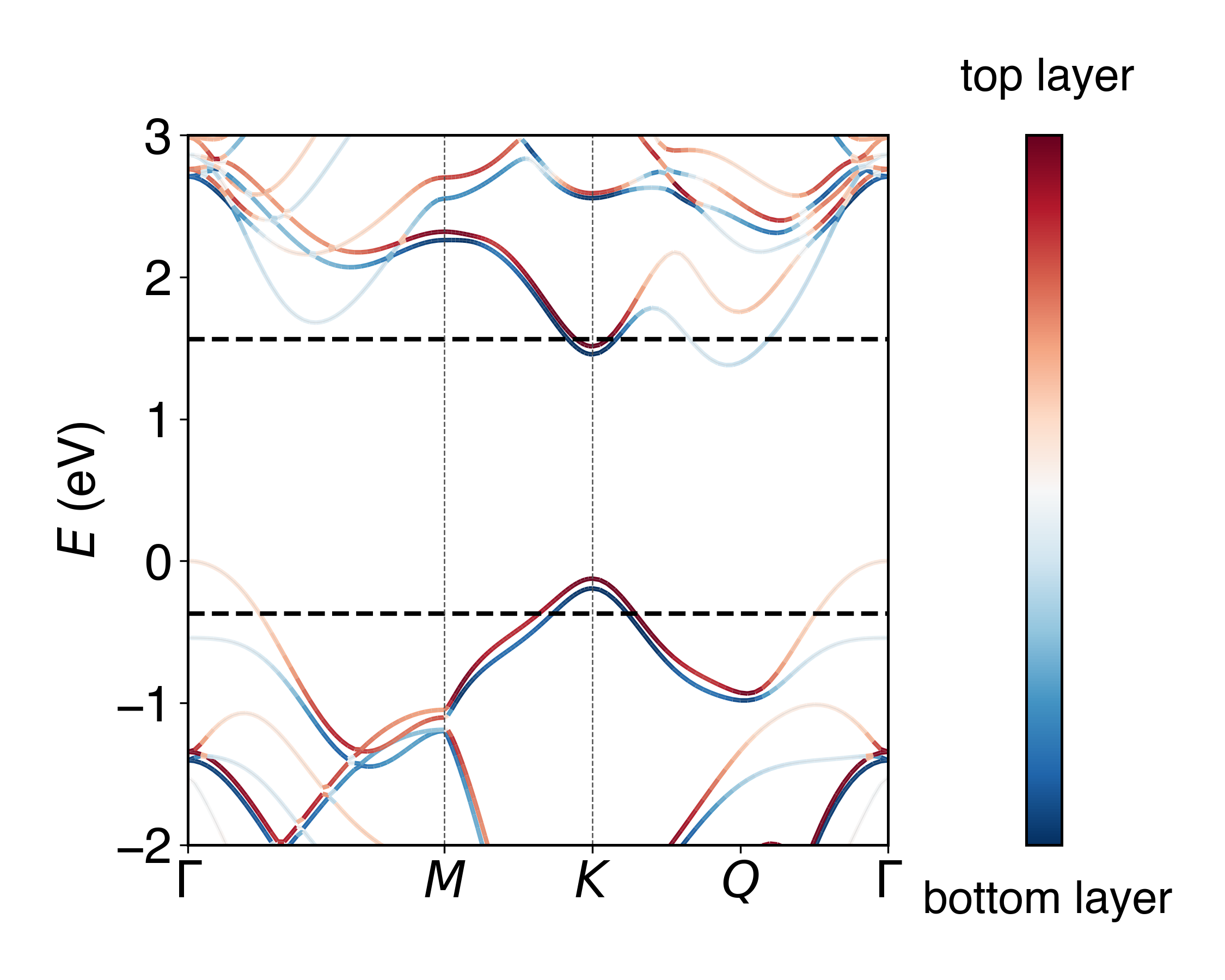}
    \caption{Band structure of WSe$_2$ projected on atoms in the top (red) and bottom (blue) layers in the ground state. The dashed lines show the two Fermi levels in cDFT calculation, for a given photoexcited carrier density n$_\mathrm{ph}$=0.3 e$^{-}$/u.c.}
    \label{fig:fig3_band}
\end{figure}

To understand the microscopic mechanism of the polarization switch, we analyze the layer-resolved electronic structure in Fig.~\ref{fig:fig3_band}, where the Kohn–Sham states are projected onto atoms in the top (red) and bottom (blue)  layers (the electronic structure of the other bilayers is reported in the SM Fig.~S1 \cite{SupplementalMaterial}).
The same stacking-induced breaking of mirror symmetry that produces the intrinsic polarization also fixes the layer character of the band-edge states.
Both the Q valley in the conduction band and the valence-band maximum at $\Gamma$ exhibit clear layer splitting that localizes the band edges on opposite layers (effective type II alignment \cite{liangOpticallyProbingAsymmetric2022}): the higher-energy valence branch is localized on the top layer, whereas the lower-energy conduction branch resides on the bottom layer. 
As a result, photoexcitation generates more holes in the top layer and more electrons in the bottom layer, producing an interlayer charge-transfer dipole.
Since the layer character of the band-edge states is fixed by the stacking configuration, the direction of this induced dipole always opposes the intrinsic polarization rather than favoring a particular absolute coordinate direction.
With increasing photoexcited carrier density, this charge redistribution strengthens and eventually overcomes the intrinsic dipole, providing a microscopic explanation for the electronic polarization reversal observed in Fig.~\ref{fig:fig1}c.

Specifically, the switching threshold is governed by the relative energies of the K and Q valleys in the conduction band. 
WSe$_2$ has the lowest Q-valley minimum among the materials studied, consistent with the stronger metal-chalcogen hybridization of the Q-valley states (with dominant in-plane $d_{xy}/d_{x^2-y^2}$ and chalcogen-$p$ character, see SM \cite{SupplementalMaterial}), and therefore allows photoexcited electrons to relax efficiently into these strongly layer-polarized states, yielding the smallest switching density. In contrast, MoS$_2$ has a much higher-energy Q valley, consistent with weaker metal-chalcogen hybridization, so electrons predominantly relax toward the nearly layer-symmetric K valley, whose conduction states are mainly $d_{z^2}$-like.
At higher excitation densities, once holes populate valence states on both layers, this imbalance is reduced, and the total polarization decreases again, consistent with the trend in Fig.~\ref{fig:fig1}c. 

Importantly, this mechanism does not require selective optical excitation of the Q valley, but rather relaxation of the photoexcited carrier distribution toward the relevant layer-polarized band-edge states. In the carrier-density regime relevant here, strong free-carrier screening supports a hot electron-hole plasma picture, so excitonic effects are not expected to qualitatively alter this relaxation pathway. The electronically driven polarization reversal is therefore governed primarily by the relaxation of the excited carriers toward these band-edge states. 

While the cDFT calculations identify the carrier densities required for electronic polarization reversal, they do not provide access to the nonequilibrium evolution of the photoexcited state or the associated time scales.
To resolve the real-time polarization dynamics, we employ the many-body approach developed in Ref.\cite{Mocatti_2026} to simulate the carrier dynamics induced by a quasi-monochromatic laser pulse of energy 2.3 eV and 30 fs duration, including explicit carrier-carrier and carrier-phonon scattering (see Section I of SM for further details \cite{SupplementalMaterial}). The optical pulse generates a photoexcited carrier density of 0.16 e$^{-}$/u.c. This density lies within the regime where cDFT predicts electronic polarization inversion and corresponds to a fluence of 1 mJ/cm$^2$, well below reported ablation thresholds in related TMD materials \cite{kim2021,solomon2022}

After the pulse, the excited carriers thermalize into a double Fermi–Dirac distribution, and we evaluate the polarization at successive time steps using the charge distributions obtained during this process. To assess the role of lattice dynamics on this timescale, we additionally included the coherent nuclear motion driven by the photoexcited carriers for $n_\mathrm{ph}=0.16$ e$^{-}$/u.c., where no dynamical instability is present (see Fig.~S4 in the SM \cite{SupplementalMaterial}). The dominant photoinduced displacements are associated with fully symmetric Raman-active $A_1$ modes. A full discussion of the photoinduced coherent nuclear motion is reported in the SM \cite{SupplementalMaterial}.

\begin{figure}[ht] 
    \centering
    \includegraphics[width=0.9\linewidth]{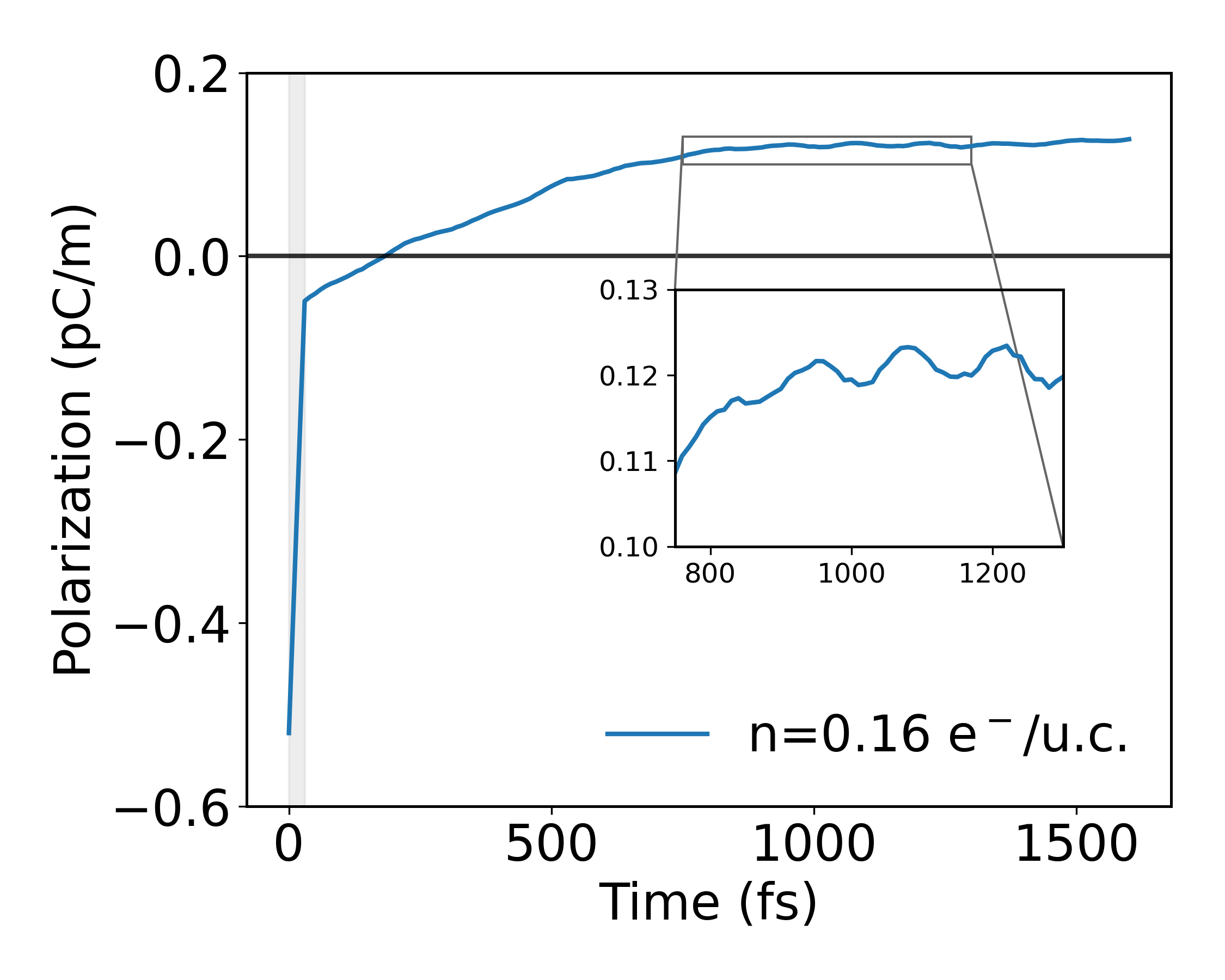}
    \caption{Time evolution of the polarization for final photoexcited carrier densities of $n_\mathrm{ph}=0.16$ e$^{-}$/u.c. (blue). The gray region (30 fs) indicates the duration of the laser pulse. 
    The boxed region is enlarged in the inset, highlighting the small oscillations from coherent phonons superimposed on the electronically driven reversal.
    }
    \label{fig:fig4_time}
\end{figure}

The resulting real-time dynamics of the out-of-plane polarization is displayed in Fig.\ref{fig:fig4_time}. This result confirms the electronic sign reversal of the total polarization identified in Fig.~\ref{fig:fig1}c. Specifically, the total polarization crosses zero at $\sim$220 fs. This time scale is much shorter than structural responses associated with phonon-mediated interlayer sliding ($\approx 10$ ps) \cite{gaoLargePhotoinducedTuning2024}. Electron-phonon coupling and the associated coherent $A_1$ modes mainly modulate the subsequent evolution of the polarization, but do not remove the sign reversal, as shown in the inset.

After the sign change, the polarization gradually relaxes to $\sim$0.12 pC/m by 1 ps, in good agreement with the cDFT results. As shown in Fig. S2 in the SM \cite{SupplementalMaterial}, the structural contribution remains subdominant relative to the electronic one, confirming that the ultrafast sign reversal is primarily electronic.

Since the effect is driven by the nonequilibrium carrier population,
the polarization is expected to retain its reversed sign while the
photoexcited charge imbalance survives and to recover its ground-state
value after carrier recombination. The timescale in related TMD systems
can exceed hundreds of picoseconds and in some cases reach the
nanosecond regime\cite{wang2019,bataller2019,karmakar2021}.

\begin{figure}[htbp]
    \centering
    \includegraphics[width=\linewidth]{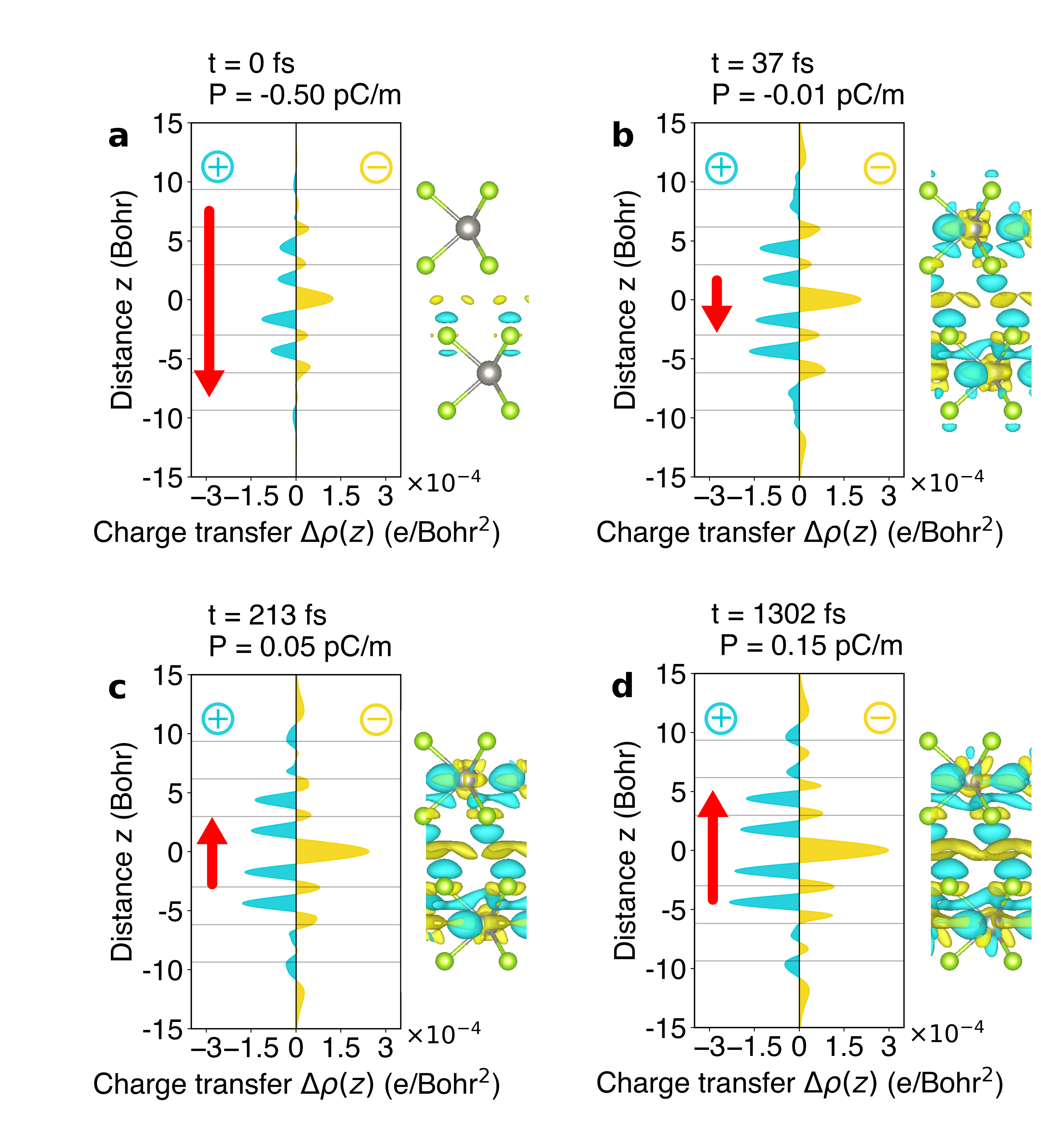}
    \caption{(a)–(d) Charge transfer at t = 0, 37, 213, and 1302 fs, respectively. Each panel shows the charge-density transfer as a function of the out-of-plane coordinate z (left) and the corresponding three-dimensional charge-transfer distribution (right), where $\Delta\rho(t)=\rho_{bl}(t)-\rho_{ml,top}-\rho_{ml,bottom}$. The cyan (yellow) corresponds to electron accumulation (depletion). The red arrow labels the sign and magnitude of the polarization at a given time $t$.}
    \label{fig:fig5_charge}
\end{figure}

To visualize the microscopic carrier dynamics responsible for switching, we isolate the charge redistribution in the bilayer relative to a simple superposition of its constituent monolayers.
Figure \ref{fig:fig5_charge} shows the time-dependent charge-density transfer, $\Delta\rho(t)=\rho_{bl}(t)-\rho_{ml,top}-\rho_{ml,bottom}$, plotted as a one-dimensional in-plane average profile along the out-of-plane $z$ direction and as a three-dimensional real-space distribution.
Here, $\rho_{bl}(t)$ is the charge density of the photoexcited bilayer at time $t$, while $\rho_{ml,top}$ and $\rho_{ml,bottom}$ are the ground-state charge densities of the isolated top and bottom monolayers.
Yellow and cyan indicate electron accumulation and depletion, respectively.
At $t=0$, Fig.~\ref{fig:fig5_charge}a shows stronger electron depletion near the top Se atoms of the bottom layer, consistent with the intrinsic polarization pointing along $-z$.
Immediately after the laser pulse, the charge distribution changes qualitatively (Fig.~\ref{fig:fig5_charge}b), most clearly around the W atoms.
In particular, electrons accumulate more strongly around W atoms in the bottom layer, producing a charge asymmetry opposite to the ground-state polarization.
This W-centered response follows from the distinct layer orbital character of the Q-valley conduction states and the $\Gamma$-point valence states shown in Fig.~\ref{fig:fig3_band}.
These processes excite electrons predominantly into bottom-layer conduction bands and deplete electronic states predominantly from top-layer valence bands, yielding the observed asymmetry.
As time evolves (Fig.~\ref{fig:fig5_charge}c,d), the asymmetry increases in magnitude, generating a vertical dipole that eventually reverses the intrinsic polarization.
The evolution of $\Delta\rho(t)$ mirrors the macroscopic polarization dynamics in Fig.~\ref{fig:fig4_time} and is consistent with the layer-selective excitation mechanism identified in Fig.~\ref{fig:fig3_band}.

\begin{figure}[t]
    \centering
    \includegraphics[width=0.70\linewidth]{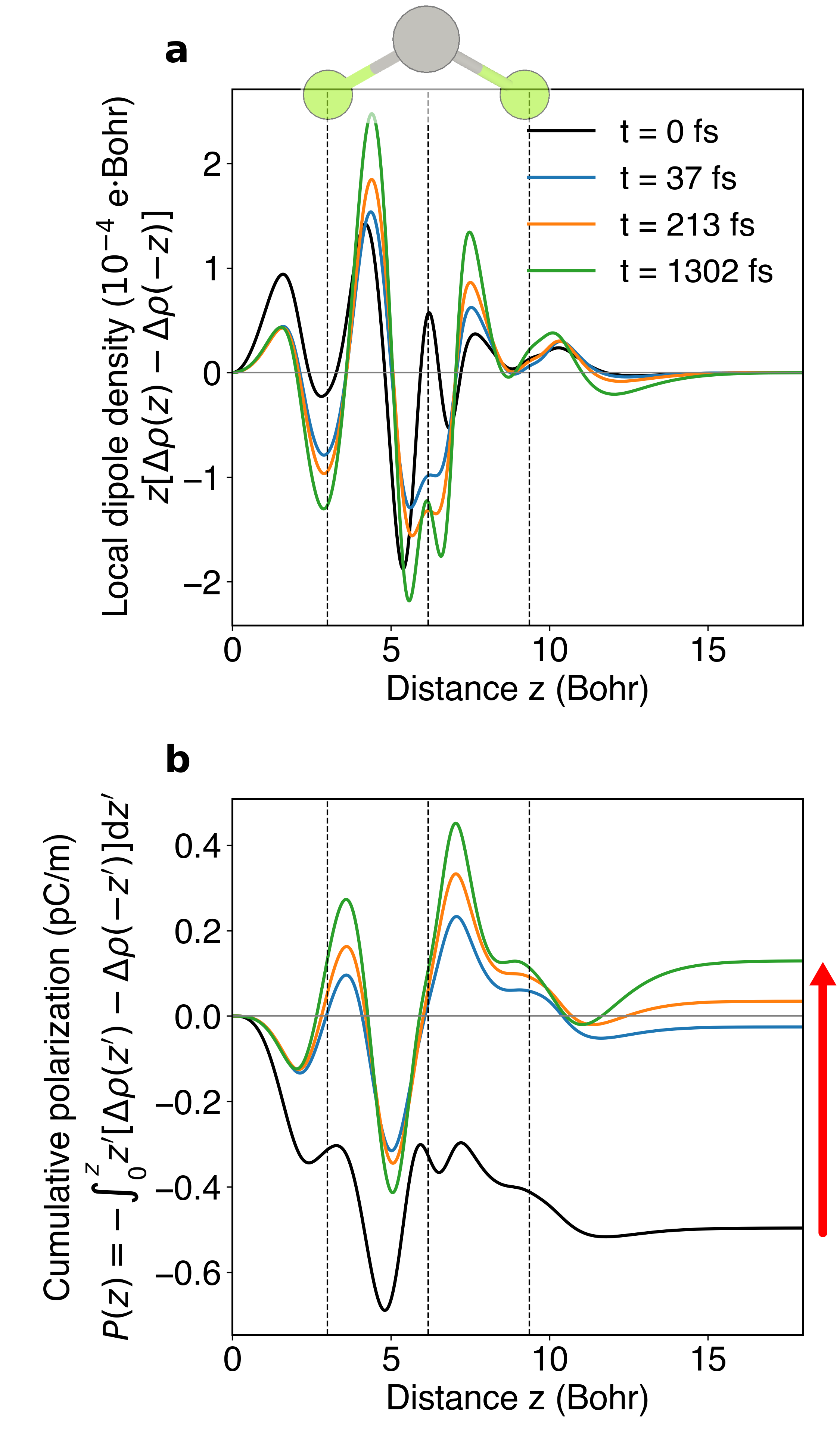}
    \caption{\textbf{a}. Local dipole density as a function of out-of-plane distance z. \textbf{b}. Cumulative polarization as a function of out-of-plane distance z. The red arrow labels the total change in polarization from $t=0$ fs to $t=1302$ fs.}
    \label{fig:fig6_localdip}
\end{figure}

To quantify how the time-dependent charge redistribution generates the macroscopic dipole, we examine the local dipole density (Fig.~\ref{fig:fig6_localdip}a) and its cumulative integral (Fig.~\ref{fig:fig6_localdip}b).
The local dipole density is defined as $z[\Delta\rho(z)-\Delta\rho(-z)]$, which extracts the antisymmetric part of the charge redistribution and weights it by the distance $z$ from the bilayer center-of-mass plane.
The prefactor $z$ converts the local charge imbalance into a position-resolved dipole moment, providing a real-space decomposition of the out-of-plane polarization.

Shortly after excitation, the local dipole density changes sign near the W atomic plane, indicating that this region is the dominant contributor to the switched polarization.
This signature directly corresponds to the W-centered electron accumulation and hole creation in Fig.~\ref{fig:fig5_charge}.
Oscillatory features also appear near the Se planes and change after excitation, providing a secondary contribution whose integrated effect is smaller than the dominant W-centered term.
To connect these local features with the total polarization, Fig.~\ref{fig:fig6_localdip}b shows the cumulative integral of the local dipole density, which tracks the spatial buildup of the macroscopic dipole.
The cumulative curve approaches a plateau at large $z$, reproducing the long-time polarization change after excitation, see Fig.~\ref{fig:fig4_time}.
The largest change occurs at the W plane, indicating that the dominant contribution to the polarization reversal originates from charge redistribution near this atomic layer along the out-of-plane direction. Because the corresponding quantity is averaged over the entire in-plane periodic unit cell, this contribution adds coherently across the crystal and contributes to the macroscopic out-of-plane polarization.
This analysis therefore provides real-space confirmation of the electronically driven mechanism underlying the ultrafast polarization reversal.

In conclusion, we demonstrate that moderate photoexcitation ($<0.2$ e$^{-}$/u.c.) can reverse the out-of-plane polarization in rhombohedral bilayer TMDs through a electronically driven mechanism, without interlayer sliding. WSe$_2$ has the lowest switching threshold, below 0.1 e$^{-}$/u.c. among the four materials examined, well below the densities required to activate phonon-driven sliding modes ($\sim$ 0.3 e$^{-}$/u.c. shown in Fig.~S4 of SM \cite{SupplementalMaterial}).
Photoexcitation generates a layer-selective charge imbalance, with electrons accumulating in the bottom layer and holes in the top, producing a transient interlayer dipole opposite to the intrinsic polarization.
Real-time many-body simulations \cite{Mocatti_2026} show that this imbalance grows rapidly and reverses the polarization within $\sim$200 fs, about 50 times faster than structural responses associated with interlayer sliding \cite{sieUltrafastSymmetrySwitch2019,fukudaUltrafastDynamicsLow2020,gaoLargePhotoinducedTuning2024}.
Coherent lattice dynamics provide only a weak periodic modulation of the total polarization and do not alter this electronically driven reversal.
Analysis of the time-dependent charge redistribution shows that the switched dipole mainly originates from a localized, W-centered charge rearrangement.
Because the process involves neither lattice reconstruction nor interlayer sliding, the polarization naturally recovers after carrier relaxation, indicating that the switching pathway is intrinsically volatile.
The same layer-selective electronic mechanism is expected to operate in a broader class of bilayer and multilayer van der Waals materials with type II layer-split band edges, suggesting that electronically driven ferroelectric control may be a general feature of this materials class. In particular, additional calculations for rhombohedral WSe$_2$ trilayer reported in the SM \cite{SupplementalMaterial} show that the same mechanism persists beyond the bilayer limit, indicating that it can extend to thicker rhombohedral stacks provided the relevant band-edge states remain layer polarized.

This ultrafast electronic switching mechanism provides a microscopic basis for using two-dimensional ferroelectrics as light-driven volatile memory devices. 
Because switching requires only moderate excitation densities and fluences, proceeds on sub-ps time scales, and resets naturally as carriers recombine, bilayer TMDs offer a promising route toward high-speed, low-energy optoelectronic memory in which information is written optically and erased automatically.
Furthermore, an electronically driven route to ferroelectric control in van der Waals materials may enable neuromorphic concepts based on short-term plasticity \cite{liShorttermSynapticPlasticity2023,zhaoUltrafastLightModulatedSliding2025}, where transient polarization states act as rapidly decaying synaptic weights. 

Experimentally, the transient electronic reversal predicted here could be detected by phase-resolved time-resolved second-harmonic generation (tr-SHG). 
Femtosecond tr-SHG has already been used to monitor ultrafast polarization
dynamics in electronic ferroelectrics, where the polar order is governed by
charge redistribution rather than by conventional ionic displacements \cite{yu2024}. 
Since conventional SHG intensity is proportional to $|\chi^{(2)}|^2$, an intensity-only measurement can reveal the suppression and recovery of the polar response but cannot by itself distinguish the two opposite signs of the polarization. 
In contrast, phase-sensitive tr-SHG measures the phase of the emitted SHG field and has been used to identify ultrafast polarization reversal through the phase of the SHG response \cite{mankowsky2017}. 
In the present case, a reversal of the out-of-plane polar contribution is expected to appear as a $\pi$ phase shift of the SHG field, providing a qualitative, sign-sensitive signature of the transient reversal of $P_z(t)$ (Further discussion see Secion VII of SM \cite{SupplementalMaterial}). 
The electronic nature of the mechanism can be tested by its fluence dependence and ultrafast onset: the SHG phase reversal should occur  within $\sim 200$ fs and below the fluence required to activate the structural instability, whereas the sliding mechanism occurs on much longer times, well above several ps.

We acknowledge fruitful discussions with Giovanni Marini.
We acknowledge the CINECA award under the Italian Super Computing
Resource Allocation (ISCRA) initiative (project IscrC\_Uf-DynFP)
for the availability of high-performance computing resources and
support. This work was funded by the European Union through the
European Research Council (ERC) project DELIGHT, No.~101052708.
Views and opinions expressed are, however, those of the author(s)
only and do not necessarily reflect those of the European Union or
the European Research Council.

The data underlying this study are openly available in the Zenodo repository at \cite{Data2026}

\nocite{
	giannozziQUANTUMESPRESSOModular2009,
	giannozziAdvancedCapabilitiesMaterials2017,
	perdewGeneralizedGradientApproximation1996,
	grimmeConsistentAccurateInitio2010,
	Hamann_2013,
	Garrity_2014,
	prandiniPrecisionEfficiencySolidstate2018,
	Monkhorst_1976,
	Marzari_1999,
	mariniLatticeDynamicsPhotoexcited2021,
	Mocatti_2026,
	Marini_2024,
	liRhombohedralstackedBilayerTransition2023,
	Trolle_2017,
	Baroni_2001,
	Dormand1980,
	Volpato_2025,
	furciFirstOrderRhombohedraltoCubicPhase2024,
	mocattiLightInducedNonthermalPhase2023,
	Virtanen_2020,
	sohierDensityFunctionalPerturbation2017,
	mommaVESTA3Threedimensional2011,
	gaoLargePhotoinducedTuning2024,
	yu2024,
	liu2025,
	mankowsky2017
}

\bibliographystyle{apsrev4-2}
\bibliography{slide_FE.bib}   
\end{document}